\documentclass[sigconf,natbib=true]{acmart}
\usepackage[ruled,linesnumbered]{algorithm2e}
\usepackage{multirow}
\usepackage{makecell}
\usepackage{enumitem}
\usepackage{diagbox}
\setitemize[1]{noitemsep,partopsep=0pt,parsep=0pt,topsep=0pt, leftmargin=10pt,
rightmargin=0pt}
\AtBeginDocument{%
  \providecommand\BibTeX{{%
    \normalfont B\kern-0.5em{\scshape i\kern-0.25em b}\kern-0.8em\TeX}}}

\setcopyright{acmcopyright}
\copyrightyear{2018}
\acmYear{2018}
\acmDOI{10.1145/1122445.1122456}

\acmConference[Woodstock '18]{Woodstock '18: ACM Symposium on Neural
  Gaze Detection}{June 03--05, 2018}{Woodstock, NY}
\acmBooktitle{Woodstock '18: ACM Symposium on Neural Gaze Detection,
  June 03--05, 2018, Woodstock, NY}
\acmPrice{15.00}
\acmISBN{978-1-4503-XXXX-X/18/06}

\begin{document}

\title{Infer As You Train: A Symmetric Paradigm of Masked Generative for Click-Through Rate Prediction}

\author{Moyu Zhang}
\affiliation{%
  \institution{Alibaba Group}
  \city{Beijing}
  \country{China}
}
\email{zhangmoyu@butp.cn}

\author{Yujun Jin}
\affiliation{%
  \institution{Alibaba Group}
  \city{Beijing}
  \country{China}
}
\email{jinyujun.jyj@alibaba-inc.com}

\author{Yun Chen}
\affiliation{%
  \institution{Alibaba Group}
  \city{Beijing}
    \country{China}
}
\email{jinuo.cy@alibaba-inc.com}

\author{Jinxin Hu}
\authornote{Corresponding Author}
\affiliation{%
  \institution{Alibaba Group}
  \city{Beijing}
  \country{China}
}
\email{jinxin.hjx@alibaba-inc.com}

\author{Yu Zhang}
\affiliation{%
  \institution{Alibaba Group}
  \city{Beijing}
  \country{China}
}
\email{daoji@alibaba-inc.com}

\author{Xiaoyi Zeng}
\affiliation{%
  \institution{Alibaba Group}
  \city{Beijing}
  \country{China}
}
\email{yuanhan@taobao.com}

\begin{abstract}
Generative models are increasingly being explored in click-through rate (CTR) prediction field to overcome the limitations of the conventional discriminative paradigm, which rely on a simple binary classification objective. However, existing generative models typically confine the generative paradigm to the training phase, primarily for representation learning. During online inference, they revert to a standard discriminative paradigm, failing to leverage their powerful generative capabilities to further improve prediction accuracy. This fundamental asymmetry between the training and inference phases prevents the generative paradigm from realizing its full potential. To address this limitation, we propose the \textbf{S}ymmetric Masked \textbf{G}enerative Paradigm for \textbf{CTR} prediction (SGCTR), a novel framework that establishes symmetry between the training and inference phases. Specifically, after acquiring generative capabilities by learning feature dependencies during training, SGCTR applies the generative capabilities during online inference to iteratively redefine the features of input samples, which mitigates the impact of noisy features and enhances prediction accuracy. Extensive experiments validate the superiority of SGCTR, demonstrating that applying the generative paradigm symmetrically across both training and inference significantly unlocks its power in CTR prediction.
\end{abstract}


\maketitle

\section{Introduction}
With the recent development of deep learning, the performance of click-through rate (CTR) prediction models in the recommender system domain has also seen rapid advancements. These models, by utilizing methods such as factorization machines (FM) \cite{fm, deepfm} and the attention mechanism \cite{hstu}, can effectively capture complex interactions among input features, thereby improving the accuracy of predictions. However, conventional CTR prediction models also face persistent challenges, such as training instability and representation collapse arising from data sparsity, which ultimately limits their generalization ability. Addressing the above challenges has therefore become a key focus of CTR prediction research filed. 

Inspired by the recent success of Large Language Models (LLMs) in the natural language processing domain, generative paradigms have gained widespread attention in recommender systems. Recommender research often treats user behavior sequences as sentences and employs auto-regressive models to predict the next item that the target user will click \cite{onerec,cobra}. While this approach moves beyond simple binary classification by introducing a generative paradigm, it fundamentally neglects crucial cross-features and item attribute features, leading to significant performance degradation \cite{dgenctr}. To address this issue, subsequent research has shifted its focus from modeling sequences to modeling the distribution of features in samples. By employing techniques such as dropout and discrete diffusion processes\cite{genctr, dgenctr}, these CTR models learn to capture complex feature dependencies. This, in turn, effectively mitigates representation collapse and enhances the modeling of feature interactions.

Despite their progress, existing generative CTR models suffer from a critical problem in their application: they employ asymmetric paradigms for training and inference. During training, the generative objectives are used to enhance the model's understanding of data distributions and to learn robust feature representations. During online inference, however, these models revert to a conventional discriminative approach, performing predictions based on potentially noisy input samples. This asymmetry means that the powerful generative capabilities acquired during training are entirely discarded at inference time, preventing these generative models from realizing their full potential in conducting predictions.

Therefore, to better leverage the potential of generative models, this paper proposes the \textbf{S}ymmetric Masked \textbf{G}enerative Paradigm for \textbf{CTR} prediction (SGCTR) which fully utilizes the model's feature generation capabilities by iteratively generating features for input samples during online inference. This allows for further improvement in prediction performance by denoising the input features, building upon the strong generalization capabilities of the network parameters acquired during training.

Specifically, during training, SGCTR employs a discrete diffusion process to predict masked features. This enables the model to learn the dependencies among features within a sample, allowing it to master plausible feature combinations. The core innovation of SGCTR lies in the inference phase, where we propose an iterative refinement strategy. In each iteration, the model first predicts the features for all masked features. Simultaneously, the prediction score for each feature is used as a confidence weight to redefine corresponding original input feature. Features with lower confidence are then re-masked for the next iteration. This process is repeated for a fixed number of steps, with the masking ratio gradually decreasing until a complete feature set is generated. At this point, all feature representations have been updated using their prediction scores, thereby leveraging the generative capabilities to denoise the model input and ultimately improve prediction performance.

The contributions of our paper can be summarized as follows:
\begin{itemize}
\item As far as we know, SGCTR is the first work to propose a symmetric generative paradigm for the training and inference phases.
\item SGCTR enhances prediction accuracy by employing an iterative refinement strategy during inference, leveraging the generative capabilities acquired during training to denoise the input. 
\item Evaluations using offline datasets and online A/B testing have demonstrated the effectiveness of SGCTR.
\end{itemize}  

\section{Preliminaries}
The objective of Click-Through Rate (CTR) prediction is to forecast the likelihood that a user will click on a specific item. Given a set of input features $\boldsymbol{X}$ and the label space $y \in \left\{0, 1 \right\}$, the CTR prediction task aims to devise a unified ranking formula $\mathcal{F}: \boldsymbol{X} \rightarrow  y$, to concurrently provide accurate and personalized prediction scores, indicating whether the target user will click the target item. The feature set $\boldsymbol{X}$ can be constructed as $[f^1, f^2, ..., f^N]$, where $N$ represents the number of the feature fields. Mathematically, the CTR prediction task estimates the probability that the target user will interact with the target item, as illustrated below:
\begin{gather}
P(y| \boldsymbol{X}) =  \mathcal{F}(f^1, f^2, ..., f^{N})
\end{gather} 

\section{Method}
While prior work has significantly improved performance by introducing generative paradigms to overcome the limitations of discriminative models, these models often merely utilize the parameters learned during training for scoring in the inference phase, failing to fully leverage the generative capabilities acquired during training \cite{genctr, dgenctr}. This paper aims to address this deficiency by proposing a novel symmetric paradigm that aligns the training and inference phases, directly applying the generative capabilities acquired during training to the online inference process.

Our approach is divided into two phases: training and online inference. As our main contribution lies in the inference phase, we primarily adopt DGenCTR \cite{dgenctr} during training to equip the model with fundamental feature generation capabilities. In the online inference phase, we introduce a novel iterative refinement strategy to redefine the model input by leveraging its generative capabilities.

\subsection{Training Phase}  
During the training phase, we employ a discrete diffusion process to learn the correlations among features. Specifically, we apply a binary mask $M \in \{0, 1\}$ to each feature field in an input sample to either preserve the feature or replace it with a mask token, representing introduced noise. The noise level is controlled by a forward process executed over discrete time steps $t$. In each training step, a time-step $t$ is sampled, and the proportion of features to be masked is determined by a noise scheduling function $\lambda(t)$, which is designed such that the mask ratio monotonically increases with $t$. 

However, the presence of high-cardinality ID features results in an intractably large output space. Calculating the full softmax over the feature spaces is computationally prohibitive. To address the issue, the sampled softmax is used to approximate the true distribution by evaluating it over a random subset of negative samples:
\begin{gather}
p(f^k|\boldsymbol{X}_{\lambda}) \approx \frac{e^{cos(f^k, G_k(\boldsymbol{X}_{\lambda})}}{\sum_{\tilde{f}^k \in S_k} e^{cos(\tilde{f}^k, G_k(\boldsymbol{X}_{\lambda}))}}
\end{gather}
where $\boldsymbol{X}_{\lambda}$ denotes the set of features that are not masked in step $t$ determined by $\lambda(t)$. $S_k$ represents the full vocabulary for the $k$-th feature field, while the denominator is approximated using ground-truth features from other instances in the batch as negative samples. The function $G_k(\cdot)$ is the generative scoring network of $k$-th feature \cite{dgenctr, hstu}. Therefore, the loss function can be defined as follows:
\begin{gather}
\mathcal{L}=\int_{0}^{1} \frac{1}{\lambda}\mathop{\mathbb{E}} [\sum -log(\frac{e^{cos(f^k, G_k(\boldsymbol{X}_{\lambda})}}{\sum_{\tilde{f}^k \in S_k} e^{cos(\tilde{f}^k, G_k(\boldsymbol{X}_{\lambda}))}}) ]d{\lambda}
\end{gather}

\subsection{Inference Phase}
As mentioned above, existing generative CTR models confine their powerful generative capabilities to the training phase. Consequently, they revert to conventional discriminative methods during online inference, relying on fixed and potentially noisy input for prediction. While they improve prediction performance by leveraging the more generalizable representations and network parameters acquired during training, they fail to fully realize the model's generative potential. Therefore, we propose a novel iterative refinement strategy designed to directly leverage the model's generative capabilities to redefine the input, thereby improving prediction accuracy.

Specifically, we establish the stable user-side features, including user profile and behavior sequences, as the initial conditions, while all other item-side and user-item cross-features are masked, $i.e. \, \boldsymbol{X}_{0}$. The iterative process then proceeds for each step $t$ as follows: \\
\emph{\textbf{$\bullet$ Feature Generation}}.  Given a masked input sample $\boldsymbol{X}_{t}$ from the current iteration step $t$, the model generates features in parallel for all masked features. For instance, the feature generated for a specific position $i$ is denoted as $\hat{f}^i_t = G_i(\boldsymbol{X}_{t})$.\\ 
\emph{\textbf{$\bullet$ Feature Confidence}}. At each masked position $i$, we calculate the similarity between the feature generated by the model at that position $\hat{f}^i_t$ and the representation of the original input feature $f^i$, using this value as the predicted score for that position, $i.e. \, c^i_t = cos(\hat{f}^i_t, f^i)$. This score is then used as the confidence score, representing the model's confidence in the prediction. For unmasked feature positions in $\boldsymbol{X}_{t}$, we directly set their confidence score to $1$. \\
\emph{\textbf{$\bullet$ Feature Redefine and Mask}}. We calculate the number of features to be masked, based on the masking scheduling function $\gamma$ by $l_t = [\gamma(\frac{t}{T})N]$, where $N$ is the total number of feature fields and $T$ is the total number of iterations. We sort the confidence scores of the features at each position from smallest to largest, mask the top $l_t$ features with the lowest confidence scores, retain the remaining features, and use the product of each feature and its corresponding confidence score as the input for the next iteration, as below:
\begin{gather}
\boldsymbol{X}_{t+1} = [f^1_{t+1}, f^2_{t+1}, ...,  f^N_{t+1}] \\
f^i_{t+1} = 
\begin{cases}
0,  & \text{if $c^i_t < {\text{Min}}[l_t]$}\\
c^i_t f^i_{t},  & \text{else}\vspace{-2mm}
\end{cases}
\end{gather}
where ${\text{Min}}[l_t]$ denotes the the top $l_t$-th lowest confidence score. 

Through this iterative redefinition process of $T$ discrete steps, we construct the final sample $\boldsymbol{X}_t$. At this point, each feature in $\boldsymbol{X}_t$ has been redefined using the model's feature generation capability. It is worth mentioning that, theoretically, if the model's generation capability is strong enough, we can replace the original features of the sample with the generated features. However, considering the high dimensionality of ID features and the limited number of training samples for the model, we will temporarily compromise by using the above redefine method.

\textbf{Prediction}. Finally, we conduct predictions as DGenCTR does. We use the iteratively redefined sample generated from the iteration process to calculate the final prediction score, as defined below:
\begin{gather}
p(y|\boldsymbol{X}_{t+1})=\frac{1}{1+e^{-(\mathcal{F}(y=1|\boldsymbol{X}_{t+1})-\mathcal{F}(y=0|\boldsymbol{X}_{t+1}))}}
\end{gather}
where the scoring function $\mathcal{F}(\cdot)$ is retrived from the training stage.

\begin{table}[t]
	\small
	\centering
	\begin{tabular}{cccc}
    		\hline Dataset &  \# Feature Fields & \# Impressions & \# Positive \\
		\hline Criteo & 39 & 45M & 26\%  \\
		Avazu& 23 & 40M & 17\% \\
		Malware & 81 & 8.9M& 50\% \\
		Industrial. & 72 & 782M & 2.8\% \\
		\hline
	\end{tabular}
	\caption{Statistics of four benchmark datasets.}
	\label{datasets}
\end{table}

\section{Experiments}
\subsection{Datasets}
To validate the efficacy of our proposed framework, four datasets are used for performance evaluation. The statistics of four datasets are detailed in Table \ref{datasets}. \textbf{$\bullet$ Criteo  \footnote{http://labs.criteo.com/downloads/download-terabyte-click-logs/}}.  It is a canonical public benchmark for CTR models, which consists of one week of real-world ad click data, featuring 13 continuous features and 26 categorical features.\textbf{$\bullet$ Avazu  \footnote{http://www.kaggle.com/c/avazu-ctr-prediction}}. It is  widely-used for CTR prediction, consisting of 10 days of chronologically ordered ad click logs with 23 feature fields in a sample. \textbf{$\bullet$ Malware  \footnote{https://www.kaggle.com/c/microsoft-malware-prediction}}. It comes from a Kaggle competition and its objective is to predict the infection probability of a Windows device.  The dataset consists of 81 feature fields. \textbf{$\bullet$ Industrial Dataset}. We gathered a dataset from an e-commerce platform's online advertising system. The training set is composed of samples from the last 20 days, while the test set is the subsequent day.

\begin{table}[t]
	\caption{Prediction performance on datasets of CTR prediction models. * indicates p-value < 0.05 in the significance test.}
	\begin{tabular}{c|c|c|c|c}
    \toprule
     \multirow{2}{*}{\diagbox{Method}{Dataset}}&  \multicolumn{4}{c}{AUC}\cr
    \cmidrule(lr){2-5}
    & {Criteo} & {Avazu} &{Malware}& {Industrial.}\cr
    \midrule
    FM &0.8037    & 0.7858  &0.7413  &0.8191      \cr
    Wide\&Deep &0.8018    & 0.7758  &0.7363  &0.8201      \cr
	DeepFM &0.8029   &0.7839  &0.7424    &0.8207    \cr
	DCN &0.8029   &0.7839  &0.7424    &0.8232    \cr
	AutoInt & 0.8025   &0.7826    &0.7403   &0.8241     \cr
FibiNET  &0.8042    &0.7832    &0.7441    &0.8250     \cr
	GDCN &0.8103  &0.7855   &0.7452   &0.8267      \cr
	MaskNet &0.8083  &0.7845    &0.7445   &0.8256     \cr
PEPNet &0.8107    &0.7860   &0.7445    &0.8272      \cr
	HSTU &0.8113    &0.7907  &0.7458   &0.8294      \cr
	\midrule 
	 HSTU+GEN &0.8142    &0.7931   &0.7479    &0.8314   \cr
	DGenCTR &0.8167  &0.7986   &0.7503   &0.8338    \cr
	\textbf{SGCTR}&\textbf{0.8186*} & \textbf{0.7999*}  & \textbf{0.7526*}  &\textbf{0.8361*}  \cr
    \bottomrule
    \end{tabular}
	\label{results}
	\vspace{-0.1cm}
\end{table}  

\subsection{Competitors}
We evaluate SGCTR against: 1) Discriminative Models: FM \cite{fm}, Wide\&Deep \cite{wdl}, DeepFM \cite{deepfm}, DCN \cite{dcn}, AutoInt \cite{autoint}, FiBiNet \cite{fibinet}, GDCN \cite{gdcn}, MaskNet \cite{masknet}, PEPNet \cite{pepnet}, and HSTU \cite{hstu}. 2) Generative Models: HSTU+GEN \cite{genctr} and DGenCTR \cite{dgenctr}. All models were implemented in TensorFlow using the Adam optimizer and Xavier initialization. The activation function was ReLU. We determined the optimal hyper-parameters through an extensive grid search. 

\subsection{Effectiveness Verification}
Table \ref{results} summarizes the overall prediction performance of all methods on all datasets, along with a statistical significance test of our model's improvements over the best-performing baseline. As shown in the table \ref{results}, our proposed method, SGCTR, consistently outperforms all models across all datasets, demonstrating its superiority.

Furthermore, our results indicate that introducing a generative paradigm enables CTR models to achieve performance competitive with that of discriminative models, showcasing the paradigm's potential. However, because these existing generative models do not apply their generative capabilities during the online inference phase, their prediction performance remains inferior to that of SGCTR. This comparison further validates the significant potential of extending the generative paradigm to the online inference phase.

\subsection{Ablation Study}
We conducted an ablation study on three variants: \textbf{SGCTR(BERT)}, which employs BERT's random masking strategy during training; \textbf{SGCTR(OneStep)}, which performs generation for all masked positions in a single step during inference; \textbf{SGCTR(GenFea)}, which directly replaces the original features with the model-generated ones during inference. The results are reported in Table \ref{result_abla}. First, SGCTR outperforms SGCTR(BERT), as the discrete diffusion process enables the model to more precisely capture the dependencies among features, which in turn enhances the feature refinement process during inference. Second, SGCTR(OneStep) also exhibits a performance drop. This is because single-step generation introduces a discrepancy between the training and inference processes, leading to less effective. Finally, SGCTR(GenFea) shows the poorest performance among all variants. This strongly suggests that, given the challenges of high-dimensional ID features, the model struggles to generate high-fidelity samples from scratch.

\begin{table}[t]
	\caption{Performance of variants on four datasets. }
	\begin{tabular}{c|c|c|c|c}
    \toprule
     \multirow{2}{*}{\diagbox{Method}{Dataset}}&  \multicolumn{4}{c}{AUC}\cr
    \cmidrule(lr){2-5}
    & {Criteo} & {Avazu} &{Malware}& {Industrial.}\cr
    \midrule
   SGCTR&\textbf{0.8186*} & \textbf{0.7999*}  & \textbf{0.7526*}  &\textbf{0.8361*}  \cr
    SGCTR(BERT) &0.8153    & 0.7958  &0.7492  &0.8326      \cr
	SGCTR(OneStep) &0.8099   &0.7897  &0.7451   &0.8281    \cr
	SGCTR(GenFea)& 0.8085   &0.7846    &0.7413   &0.8215     \cr
    \bottomrule
    \end{tabular}
	\label{result_abla}
	\vspace{-0.1cm}
\end{table}

\subsection{Hyper-Parameters Sensitivity Analysis}
\begin{figure}[t]
  \centering
  \includegraphics[width=\linewidth]{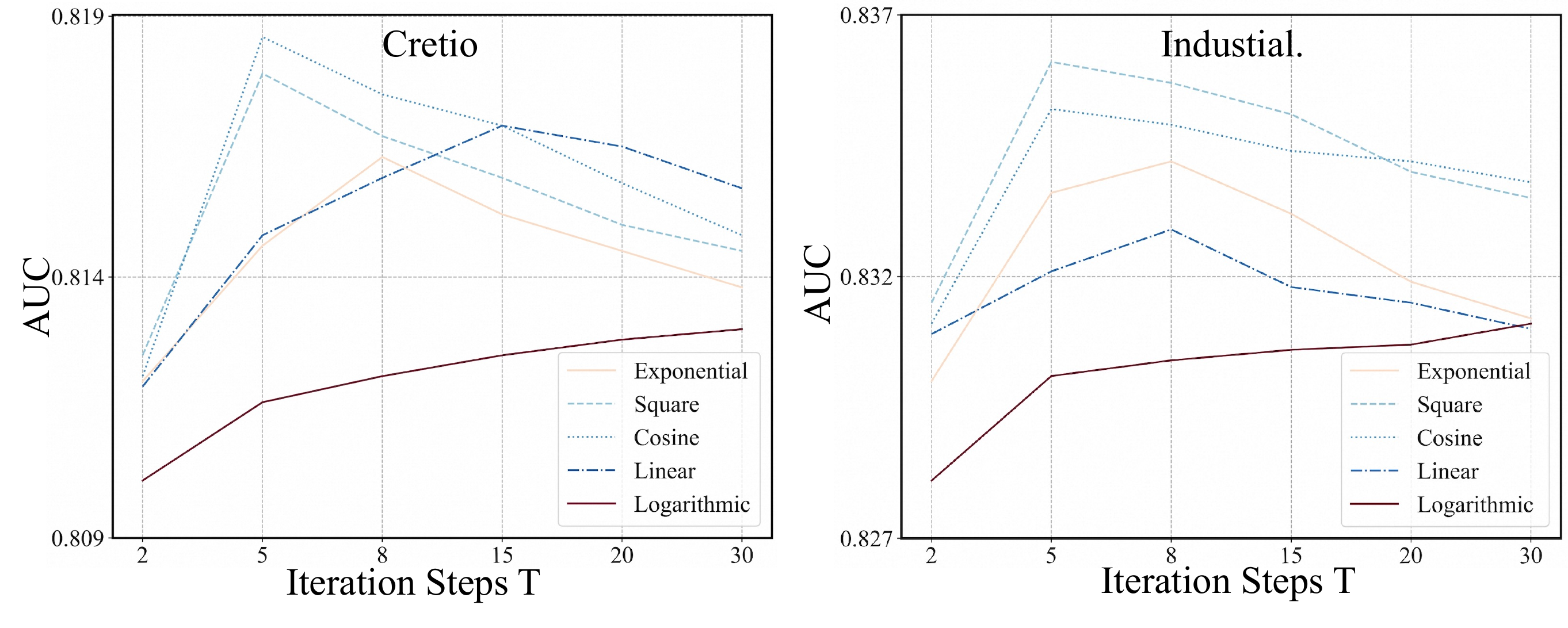}
  \caption{Prediction performance under different steps.}
  \label{abla}
\end{figure}

In the inference phase of SGCTR, the mask scheduling function $\gamma$ and iteration steps $T$ will influence the model's performance. We evaluated the effectiveness of five schedules: exponential, square, cosine, linear, and logarithmic and observed that the cosine and square schedules performed best across all datasets. This result is intuitive: as concave functions, they ensure a high mask ratio in the early iterations, compelling the model to focus on refining a small number of core features first. As the iterations progress, the number of redefined features gradually increases. In contrast, the exponential schedule, which attempts to redefine a large number of features in the final steps, resulted in performance degradation. Furthermore, we experimented with varying the number of iteration steps, $T$ with the results shown in Figure \ref{abla}. We observed that as $T$ increases, performance initially improves, reaches an optimal point, and then begins to decline. This indicates that an excessive number of iterations is counterproductive, as it would impose overly strict constraints on the feature selection at each step and place excessively high demands on the model's prediction scores.

\begin{table}[t]
	\caption{Performance of different functions on four datasets. }
	\begin{tabular}{c|c|c|c|c}
    \toprule
     \multirow{2}{*}{\diagbox{Function}{Dataset}}&  \multicolumn{4}{c}{AUC}\cr
    \cmidrule(lr){2-5}
    & {Criteo} & {Avazu} &{Malware}& {Industrial.}\cr
    \midrule
    Exponential  &0.8163    & 0.7988  &0.7508  &0.8342     \cr
    Square &0.8179    & \textbf{0.7999*}  &0.7522  &\textbf{0.8361*}      \cr
    Cosine &\textbf{0.8186*} & 0.7991  & \textbf{0.7526*}  &0.8352  \cr
	Linear&0.8169  &0.7979  &0.7502   &0.8329    \cr
	Logarithmic& 0.8130   &0.7934    &0.7481   &0.8311     \cr
    \bottomrule
    \end{tabular}
	\label{result_function}
	\vspace{-0.1cm}
\end{table}

\subsection{Online A/B Testing Results}
We conducted a ten-day online A/B test from October 16–26, 2025, on a large-scale e-commerce platform. SGCTR was tested against a production baseline model that also employs a DGenCTR-based architecture. The results showed that SGCTR achieved a significant 2.1\% relative improvement in click-through rate (CTR), strongly demonstrating the practical value and effectiveness of SGCTR.

\textbf{Online Deployment}. For the online deployment, an iteration step of $T=5$ provides the optimal trade-off between performance and latency. To mitigate the computational overhead, we cache the keys and values after the initial computation. Consequently, subsequent iterations only need to re-compute the query and ensures that the overall computational complexity remains at $O(n)$.

\section{Conclusions}
In this paper, we propose the Symmetric Masked Generative Paradigm for CTR Prediction (SGCTR), a novel framework that establishes symmetry between the training and inference phases. Specifically, after learning feature dependencies to build its generative capabilities during training, SGCTR leverages these capabilities during inference to iteratively redefine the input sample's features. This process mitigates the impact of noisy features and significantly improves the model's prediction accuracy. Finally, extensive experimental results demonstrate that SGCTR effectively unleashes the full potential of generative paradigms in CTR prediction models.

\end{document}